\documentclass[aoas,nameyear,dvips]{arximspdf}
\usepackage{graphics}

\doi{10.1214/09-AOAS269}
\volume{3}
\issue{4}
\pubyear{2009}
\firstpage{1695}
\lastpage{1709}

\begin{document}
\begin{frontmatter}

\title{Bayesian inference and model choice in a hidden stochastic
two-compartment model of hematopoietic stem cell fate decisions\protect\thanksref{T1}}
\runtitle{Bayesian inference for HSC fate decisions}

\begin{aug}
\author[a]{\fnms{Youyi} \snm{Fong}\ead[label=e1]{yfong@u.washington.edu}},
\author[b]{\fnms{Peter} \snm{Guttorp}\ead[label=e2]{peter@stat.washington.edu}\corref{}}
\and
\author[c]{\fnms{Janis} \snm{Abkowitz}\ead[label=e3]{janabk@u.washington.edu}}
\runauthor{Y. Fong, P. Guttorp and J. Abkowitz}
\affiliation{University of Washington}
\address[a]{Y. Fong\\
Department of Biostatistics\\
University of Washington\\
Seattle, Washington 98005\\
USA\\
\printead{e1}}
\address[b]{P. Guttorp\\
Department of Statistics\\
University of Washington\\
Seattle, Washington 98005\\
USA\\
\printead{e2}}
\address[c]{J. Abkowitz\\
School of Medicine\\
University of Washington\\
Seattle, Washington 98005\\
USA\\
\printead{e3}}
\thankstext{T1}{Supported by the National Institutes of Health Grants
R01-HL46598 and R01-HL082933.}
\end{aug}

\received{\smonth{1} \syear{2009}}
\revised{\smonth{6} \syear{2009}}

%
\begin{abstract}
Despite rapid advances in experimental cell biology, the in vivo
behavior of hematopoietic stem cells (HSC) cannot be directly observed
and measured. Previously we modeled feline hematopoiesis using a
two-compartment hidden Markov process that had birth and emigration
events in the first compartment. Here we perform Bayesian statistical
inference on models which contain two additional events in the first
compartment in order to determine if HSC fate decisions are linked to
cell division or occur independently. Pareto Optimal Model Assessment
approach is used to cross check the estimates from Bayesian inference.
Our results show that HSC must divide symmetrically (i.e., produce two
HSC daughter cells) in order to maintain hematopoiesis. We then
demonstrate that the augmented model that adds asymmetric division
events provides a better fit to the competitive transplantation data,
and we thus provide evidence that HSC fate determination in vivo occurs
both in association with cell division and at a separate point in time.
Last we show that assuming each cat has a unique set of parameters
leads to either a significant decrease or a nonsignificant increase in
model fit, suggesting that the kinetic parameters for HSC are not
unique attributes of individual animals, but shared within a species.
\end{abstract}

%

\end{frontmatter}

\section{The biology of hematopoiesis}\label{sec_biology}

Hematopoiesis is the process of blood cell production. More precisely, it is
the process in which hematopoietic stem cells (HSCs) make fate decisions and
through sequential divisions, differentiate into progenitor cells. These cells
in turn differentiate into white blood cells, red blood cells or platelets.
While a lot is known about how progenitor cells differentiate, since their
behavior has been studied both in vivo and in vitro, relatively little is
known about HSC. This is due to the fact that HSCs are difficult to isolate,
as they do not have a completely unique physical or antigenic phenotype. In
addition, in vivo, HSC decisions depend on input from neighboring cells and
cytokines and not just their intrinsic cell programming. HSCs support the
entire blood and immune system, and can reconstitute hematopoiesis after
transplantation. Understanding their kinetics is of great importance. For
example, this could lead to new treatments for leukemia and more effective
clinical HSC transplantation procedures.

An HSC basically has to fulfill two directives, to self-renew and to
differentiate. In addition, like all cells, an HSC eventually will die through
apoptosis. An HSC self-renews by dividing symmetrically into two identical
daughter cells, each of which becomes a new HSC. Since the normal function of
an HSC requires input from cells and cytokines in its bone marrow
microenvironment, termed niche, a newly born HSC will die unless it finds a
niche in time. Biological data argue that there is a limited number of niches
available in an organism and this helps maintain the number of HSC in a steady
state in a normal adult [\citet{abkowitz2002enh}, \citet{czechowicz2007etv}].
The pool of progenitor cells is replenished by the differentiation of HSC,
which can happen in two ways conceptually. First, an HSC can be cued to commit
to a specific progenitor fate. Alternatively, an HSC can divide into two
cells, which are fated to become an HSC or a progenitor cell respectively at
or right after the time of division. We call this ``asymmetric
division,'' although from a mathematical perspective, one
cannot distinguish a fate decision programmed at mitosis from a fate decision
resulting from microenvironmental input at or immediately following cell division.

In Drosophila germ line cells, it is clear that the cell division event is
indeed asymmetric, that is, fate depends on the spindle orientation relative to
the Hub cell (niche) and results from the unequal distribution of
intracellular regulators and extracellular (Hub-derived) signals between
daughter cells during mitosis [reviewed in
\citet{Knoblich2008}
and
\citet{yamashita2008}]. In the mammalian system, this is less certain, although studies of murine
neuroprogenitors, muscle satellite cells and T cells (following contact with
an antigen presenting cell) suggest this occurs [\citet{Knoblich2008} and
\citet{chang2007}]. The elegant studies of Wu and colleagues
[\citet{wu2007ihp}] show that murine HSC/progenitor cells, defined as immature
by virtue of Notch transcription, divide asymmetrically in vitro. There is no
in vivo evidence to suggest that asymmetric divisions happen, however, and
importantly, observing asymmetric outcomes does not require that HSC division
and fate determination are mechanistically linked [see discussion in
\citet{Schroeder2007}].

In order to get an idea of the contributions of the feline hematopoietic stem
cells to the progenitor cells, a specific set of experiments was designed
using female Safari cats. Safari cats are the offspring of matings between a
domestic cat and the South American Geoffroy wild cat. Since these two breeds
of cat have had long separate developments, they express an
electrophoretically distinct phenotype of the \mbox{X-chromosome-linked} enzyme
glucose-6-phosphate dehydrogenase (G6PD). During embryogenesis, since either
the paternal or the maternal \mbox{X-chromosome} is inactivated, the female Safari
cats have some somatic cells expressing the domestic-type G6PD (d G6PD) and
other expressing the Geoffroy-type G6PD \mbox{(G G6PD)}. The G6PD phenotype is
retained after replication and differentiation, and it is functionally
neutral. That is, the cells that express it do not have significant
self-renewal or differentiation advantages. Therefore, it provides a binary
marker or label of each cell and its offspring. For more details on the
experiment, see
\citeauthor{abkowitz1988} (\citeyear{abkowitz1988,abkowitz1990,abkowitz1993}).

Observing the percentage of progenitor cells expressing the d G6PD phenotype
over a period of almost 6 years in normal female Safari cats (with
observations taken approximately every 4 weeks) did not seem to provide much
information about the HSC behavior. In fact, this percentage remained
relatively constant during the six years of observation, suggesting that
hematopoiesis is a polyclonal and stable process.

Since there might be more information in observing the hematopoiesis process when it is supported by a much smaller number of stem cells, a
number of female Safari cats were irradiated in order to kill their bone
marrow (where HSCs reside in adults) and a small number of bone marrow cells,
collected prior to the radiation, were transplanted back. Since large animals
have relatively few HSCs compared to other cells in the marrow, at the start
of the experiment the transplanted cats are likely to contain a very small
number of HSCs. For this reason the process modeling the HSC behavior should
have a discrete state-space rather than a continuous one. The behavior of the
binary label (d G6PD versus G G6PD) within the progenitor cells was then
monitored in samples taken every two to six weeks. Under this setting, the
percentage of labeled cells is more variable over time. For example, some cats
showed wide clonal fluctuations during the first year or so and then
stabilized, suggesting that initial hematopoiesis after the transplantation
was supported by only one or two clones.

\section{A stochastic model of hematopoiesis}\label{sec_model}

\begin{figure}[b]

\includegraphics{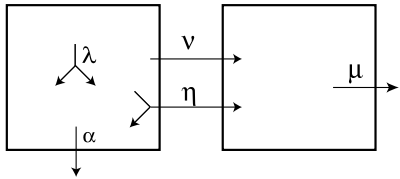}

\caption{A stochastic model of hematopoiesis.}%
\label{bedaa2}%
\end{figure}

We model the hematopoiesis as a two-compartment model. The first compartment
contains the HSC, while the second contains the clones (i.e., the entire
production) of committed stem cells. In the first compartment we allow
stochastic decisions to divide, specialize or die, with fates depending on the
type of event. Thus, an HSC follows a birth (symmetric division with rate
$\lambda$; when the maximum number of niches is reached, one daughter cell
dies immediately), death (apoptosis at rate $\alpha$) and emigration (through
commitment with rate $\nu$ or through asymmetric division with rate $\eta$).
The second compartment gets input from committed stem cells, and the duration
of productive life of a clone is modeled exponentially (clonal death with rate
$\mu$). Figure \ref{bedaa2} is a schematic description of the stochastic
model. Table \ref{TabNotation} lists all the stochastic events.%

\begin{table}
\caption{A list of stochastic events. All events are modeled as Poisson
events and the rate parameters are listed in the last column}\label{TabNotation}
\begin{tabular*}{\textwidth}{@{\extracolsep{\fill}}lcc@{}}
\hline
\textbf{Name} & \textbf{Short name} & \textbf{Rate parameter}\\
\hline
Symmetric division & S & $\lambda$\\
Commitment & C & $\nu$\\
Clonal death & D & $\mu$\\
Asymmetric division & As & $\eta$\\
Apoptosis & Ap & $\alpha$\\
\hline
\end{tabular*}
\end{table}

In this paper we will be studying several submodels of the one shown in
Figure \ref{bedaa2} and a convention is adopted to name each model by
concatenating the short names of the stochastic events it contains. For example,
a model containing the symmetric division event, the commitment event and the
clonal death event is called a SCD model; add the asymmetric event and the
apoptosis event and we get the full model SCDAsAp.

It follows from the description of the experiment above that the total number
of cells both in compartments one and two in Figure \ref{bedaa2} can be seen as
the sum of two independent and identically distributed population processes
that differ only in a label: d G6PD or G G6PD. In short, the population process
is two-dimensional in both compartments. One dimension is the population of
cells expressing the d phenotype, the other is the population of cells
expressing the G phenotype.

Denote the rate parameters collectively as $\rho$. Denote by $\omega$ the set
of events between time 0 and time $T$ or, equivalently, the number of cells
over time in that time span. We call $\omega$ the path or state. Suppose
$\omega$ is composed of $n$ events, which divide the time span $[
0,T]  $ into $n+1$ intervals denoted by $t_{0},\ldots,t_{n}$. The
probability density of $\omega$ given the rate parameters $\rho$ is%
\begin{eqnarray*}
f(  \omega|\rho)   &  =&\prod_{k=0}^{n-1}m_{k}\exp[  -(
z_{k}\lambda+z_{k}\nu+z_{k}\alpha+z_{k}\eta+x_{k}\mu)  t_{k}] \\
& &{} \times\exp[  -(
z_{n}\lambda+z_{n}\nu+z_{n}\alpha+z_{n}\eta+x_{n}\mu)  t_{n}]  ,
\end{eqnarray*}
where $z_{k}$ and $x_{k}$ are the numbers of cells/clones (either domestic or
Geoffroy) in the first and second compartments, respectively, at the time of
 the $k$th event ($z_{0}$ and $x_{0}$ are the numbers at time 0), $m_{k}%
=z_{k}\lambda$, $z_{k}\nu$, $z_{k}\alpha$, $z_{k}\eta$ or $x_{k}\mu$
depending on whether the $k$th event is a symmetric division, a commitment,
an apoptosis, an asymmetric division event in the first compartment or a
death event in the second compartment. Note that $z_{k}$ and $x_{k}$ are not
observed. We choose $z_{0}$ and $x_{0}$ to be 10 and 5, respectively (either
domestic or Geoffroy) via a combination of prior belief and sensitivity study
[\citet{golinelli2006bih}].

In order to relate this model of the unobserved HSCs and their progenitors to
the actual observations, we assume that each clone contributes equally to
hematopoiesis, and sample certain types of cells (BFU-E plus CFU-GM). By the
assumption made, the number of d-labeled cells in the samples should be
binomially distributed with success probability reflecting the proportion of d
HSC in the second compartment. Specifically, denoting $t_{i}$ the time of the
$i$th observation, $N(  t_{i})  $ the number of progenitor cells
in the sample at time $t_{i}$, $x_{d}(  t_{i})  $ and $x_{G}(
t_{i})  $ the numbers of d-type and G-type clones in the second
compartment at time $t_{i}$, the~distribution of the number of d-type cells
$Y(  t_{i})  $ is%
\[
[  Y(  t_{i})  |\omega]  \sim\operatorname{Binomial}\biggl(
N(  t_{i}),\frac{x_{d}(  t_{i})  }{x_{d}(
t_{i})  +x_{G}(  t_{i})  }\biggr)  .
\]

\section{Some history}\label{sec_history}

From a statistical point of view, parameter estimation needs to involve the
likelihood function. In the first set of papers in the long-standing
collaboration between Guttorp and Abkowitz [%
\citet{abkowitz1990}, \citet{guttorp1990smh}%
] we were able to perform a complete likelihood analysis, with a nonstandard
shape of the likelihood due to the initial number of transplanted HSC being a
parameter. There were serious numerical difficulties in evaluating the
likelihood, and a Markov chain Monte Carlo approach to inference with varying
state space was developed to do a Bayesian analysis of the early cat
transplant data [%
\citet{newton1992bis}%
].

When additional cat data proved this early model a poor fit, we developed the
first version of the current model, which initially was analyzed using
simulation tools. Using some features of the observed data, we developed
objective criteria to assess simulated paths. When a set of simulated paths
failed one or more of the criteria, those parameter values were deemed
infeasible. In essence, we were using the multiple decision tools of Pareto
Optimization [%
\citet{vincent1981ops}%
] and the Pareto Optimal Model Assessment approach by
\citet{reynolds1999mca}
to determine possible parameter values, with an ad hoc approach to find
``optimal'' parameter values.

A statistical approach, using estimating equations, was developed by Catlin in
her dissertation [\citet{catlin1997sip}] and applied to the cat data [%
\citet{catlin2001sit}%
], yielding parameter estimates with a tighter range than those obtained from
the Pareto optimization.
\citet{golinelli2000bih}
studied a simplified version of the model and developed tools for calculating
the posterior distribution using Markov chain Monte Carlo methods. It was
computationally infeasible at the time to run sufficient numbers of MCMC
iterations for the convergence of the algorithm. With improved computing
facilities and a faster algorithm,
\citet{golinelli2006bih}
then calculated the posterior distributions for one of the first transplanted
cats, as well as that using all the cats. These distributions agree with and
thus confirm our previous simulation-based work, and are more precise than the
alternative method developed by
\citet{catlin2001sit}%
.

In this paper we extend the Bayesian inference machinery developed in
\citet{golinelli2006bih}
 for the SCD model to work with other models. Basically we draw from the
posterior distribution $\rho,\omega|Y$ by using a Gibbs sampler that
alternates between updates of $\rho|\omega,Y$ (parameter update) and
$\omega|\rho,Y$ (state update). The parameter update is rather easy with the
choice of gamma or uniform prior because we can write down the conditional
distribution of $\rho|\omega,Y$ analytically. The state update is achieved by
a Reversible Jump Metropolis--Hastings algorithm [%
\citet{robert2004mcs}%
]. Details are given in the \hyperref[append]{Appendix}.

The increased complexity in both the parameter space and the state space
requires that we run more MCMC iterations than before and the time to run it
is getting close to being prohibitive. In order to improve the performance, we
allow the state update for each cat to run in its own thread, thus speeding up
state updates, the more time-consuming part of the algorithm. The program was
run on a custom-built high-performance workstation. This allows us to run more
complicated models, as well as simultaneously analyze all the data from the
transplanted cats. In particular, we are able to fit various refined models
(Sections \ref{sec_essential} and \ref{sec_bedas}) and assess the hypothesis that
all cats have the same parameter values (Section~\ref{sec_homog}). We still
use the Pareto Optimal Model Assessment approach to rule out certain models
(Section \ref{sec_essential}).

\section{Essentialness of apoptosis, symmetric division
and commitment events}\label{sec_essential}

Based on scientific evidence we know that apoptosis does happen to all somatic
cells.
\citet{golinelli2006bih}
chose to leave out this event because simulation studies done in
\citet{abkowitz1996ehm}
indicated that apoptosis was not essential in reconstructing our experimental
observations and because of the computational challenge in implementing it.
With the recent progress in software and hardware, we attempt reversible jump
MCMC for the \mbox{SCDAp} model. The posterior means for $\lambda,\nu,\mu,\alpha$ are
0.084, 0.00019, 0.038, 0.0024, separately. To find out whether the SCDAp model
provides a better fit to the data than the SCD model, we estimate Bayes factors.
To this end, we estimate the integrated likelihood under each model with a
stabilized harmonic mean estimator [%
\citet{raftery2007eil}%
]%
\[
\hat{p}_{\theta}(  y|M)  =\Biggl(  \frac{1}{B}\sum_{i=1}^{B}\frac
{1}{\pi(  y|\theta_{i},M)  }\Biggr)  ^{-1},
\]
where $\theta$ is any subset of parameters in model $M$, $\theta_{i}$ is one
draw from the posterior distribution of $\theta$, $\pi(  y|\theta
_{i},M)  $ is the density of the data under model $M$ with all
parameters other than $\theta$ integrated out, and $B$ is the number of draws
used to calculate the harmonic mean. Depending on the choice of $\theta$, the
estimator can have large or small variance. In our model, if we choose
$\theta$ to be the path, then the estimator has very large variance (data not
shown). But if we let $\theta$ be either $\lambda$, $\nu$ or~$\mu$, then
$\hat{p}_{\lambda}$, $\hat{p}_{\nu}$ and $\hat{p}_{\mu}$ all have much
smaller variance and are similar to each other (Table \ref{TabIntLik}). The
Bayes factor for comparing the \mbox{SCDAp} model and the SCD model, $\hat{p}%
_{\theta}(  y|\mathit{SCDAp})/\hat{p}_{\theta}(  y|\mathit{SCD})  =0$,
indicates that the SCD model is strongly preferred over the \mbox{SCDAp} model for
describing this data set. Since SCD is a submodel of \mbox{SCDAp}, this result means
that without the presence of the apoptosis event, the SCD model does a good job of
describing the data set. Adding the  apoptosis event results in a more complex
model, which is penalized by the Bayes factor approach. This result is
consistent with
\citet{abkowitz1996ehm}
which shows by the Pareto Optimal Model Assessment approach that the data
contain little information about the apoptosis rate.

\begin{table}
\caption{Integrated likelihoods of cats for various models. 100 weeks data.
Gamma priors are assumed}\label{TabIntLik}
\begin{tabular*}{\textwidth}{@{\extracolsep{\fill}}lccccc@{}}
\hline
\textbf{Models} & $\bolds{\log\hat{p}_{H}(y) } $ & $\bolds{\log\hat{p}_{\theta_{1}}(y) } $ & $\bolds{\log\hat{p}_{\theta_{2}}(y)}$ &$\bolds{\log\hat{p}_{\theta_{3}}(y)}$ & \textbf{BF}\\
\hline
SCD & $-$682.15 & $-$647.27 & $-$649.24 & $-$646.35 &\phantom{0}1.0\phantom{0}\\
SCDAp & $-$692.04 & $-$659.28 & $-$662.61 & $-$660.09 &\phantom{0}0.00\\
CDAsAp& $-$673.11 & $-$645.84 & $-$649.01 & $-$647.41 &\phantom{0}1.2\phantom{0}\\
SDAsAp & $-$689.72 & $-$668.17 & $-$672.57 & $-$676.71 &\phantom{0}0.00\\
SCDAs & $-$670.34 & $-$644.59 & $-$644.30 & $-$645.16 &19\phantom{.00}\\
\hline
\end{tabular*}
\end{table}

Since asymmetric division can be viewed as combining symmetric division and
commitment in one step, questions arise as to whether it is possible to leave
out either symmetric division or commitment in the presence of asymmetric
division. To check the essentialness of symmetric division, we performed a
Bayesian analysis of the CDAsAp model using a Gamma prior. The Bayes factor
comparing the CDAsAp model with the SCD model is $1.23$ (Table \ref{TabIntLik}%
). This Bayes factor is nonconclusive. We simulate 50 virtual cats from the
CDAsAp model using the means of the posterior distributions of the rate
parameters ($\nu=0.00738$, $\mu=0.05969$, $\eta=0.03338$, $\alpha
=0.00426$). We then assess each virtual cat with the five assessment criteria
developed in
\citet{abkowitz1996ehm}%
. The distributions of each of the first four criteria in the observed and in
the virtual cats are compared using Kolmogorov--Smirnov tests and $P$ values are
0.00, 0.000, 0.000 and 0.000. The die-out rate of the virtual cats is 100\%.
We also simulate virtual cats from six randomly chosen sets of parameters from
the posterior distributions. While the first four tests produce variable $P$ values, the die-out rates are all higher than 96\%.
This strongly suggests that symmetric division is required to explain the
data, that is, the hematopoietic reserve (compartment 1) cannot regenerate
and cannot maintain continued hematopoiesis unless HSC symmetrically divide.

Similarly, we did a Bayesian analysis of the SDAsAp model using a Gamma prior.
The Bayes factor comparing the SDAsAp model with the SCD model is~$0.00$
(Table \ref{TabIntLik}), suggesting that commitment is required to explain the
data. We simulate 50 virtual cats from the SDAsAp model using the means of the
posterior distributions of the rate parameters ($\lambda=0.0659$%
, $\mu=0.04538$, $\eta=0.00136$, $\alpha=0.00142$). Comparison of the first
four criteria between the virtual cats and the observed cats results in the
$P$
values of 0.073, 0.005, 0.000 and 0.75. The die-out rate of the virtual cats
is 0\%. We also simulate virtual cats from six randomly chosen sets of
parameters from the posterior distributions. The $P$ values from testing the
third criterion, the relative amount of variation in the d-type progenitor
percentage immediately following transplantation [%
\citet{abkowitz1996ehm}%
], are all 0.000. This further supports the notion that commitment is
essential to HSC kinetics. The maintenance of hematopoiesis by a persisting
cohort of progenitor cells is not compatible with experimental observations.

\section{Bayesian inference for SCDAs model}\label{sec_bedas}

Table \ref{TabBEDAs} reports the means and 95\% HPD credible intervals of the
posterior distributions of the parameters for the SCDAs model. Those for the
SCD model are reported as well for comparison purpose. For the SCDAs model,
the best estimates for $\lambda$, $\nu$, $\mu$ and $\eta$ are 1 per 11 weeks,
1~per 13 weeks, 1 per 5.2 weeks, and 1~per 13 weeks, respectively. The
estimated symmetric division rates and commitment rates are largely similar
between the two models, while the clonal death rate in the SCDAs model is a
bit larger than that in the SCD model. This makes sense because in the SCDAs
model, in addition to commitment events, asymmetric division events also
contribute to the increase in the number of clones in the second compartment,
thus requiring an increase in the death rate to maintain a steady state.

\begin{table}[b]
\caption{The means and 95\% HPD credible intervals of the posterior
distributions of parameters for SCDAs and SCD models. Gamma prior and 100
weeks data}\label{TabBEDAs}
\begin{tabular*}{\textwidth}{@{\extracolsep{\fill}}l@{\ \ }c@{\ \ }c@{\ \ }c@{\ \ }c@{}}
\hline
& \textbf{Symm division} $\bolds{(\lambda)}$ & \textbf{Commitment} $\bolds{(\nu)}$ & \textbf{Clonal death} $\bolds{(\mu)}$ & \textbf{Asym. division} $\bolds{(\eta)}$\\
\hline
SCDAs & 0.093 (0.044, 0.141) & 0.079 (0.036, 0.125) & 0.193 (0.118, 0.271) &0.078 (0.009, 0.159)\\
SCD & 0.094 (0.052, 0.148) & 0.079 (0.038, 0.137) & 0.139 (0.081, 0.209) &\\
\hline
\end{tabular*}
\end{table}

\begin{figure}

\includegraphics{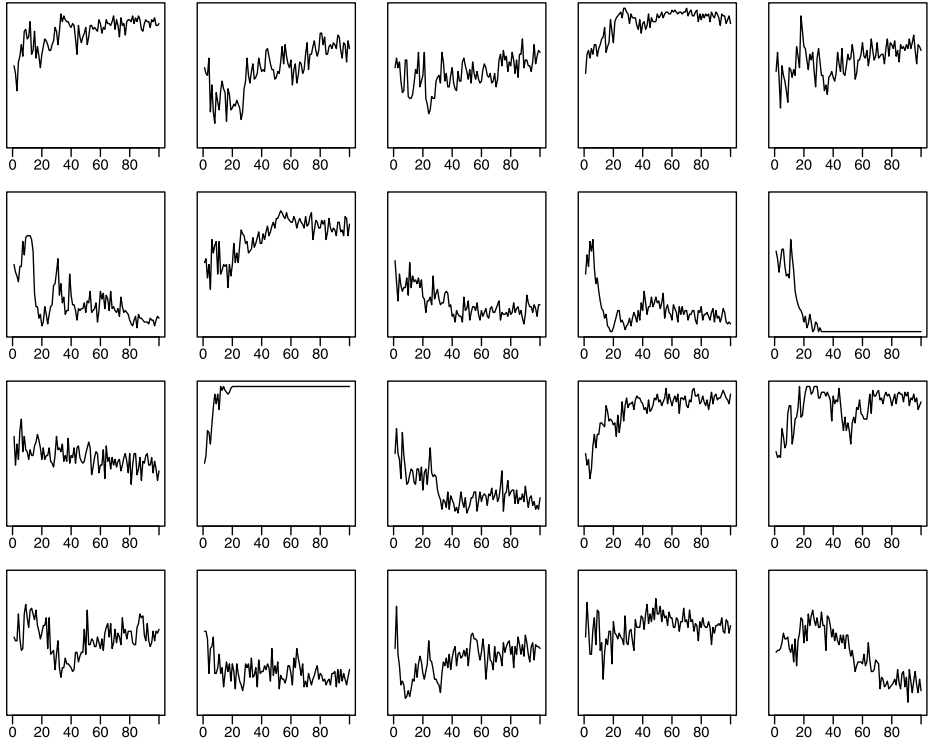}

\caption{Cats simulated from the mean of the posterior distributions of the
SCDAs model. Gamma priors are assumed.}
\label{bedas_sim.eps}
\end{figure}

The Pareto Optimal Model Assessment approach was previously used in
\citet{abkowitz1988}
to find the acceptable range of the parameters in the SCDAp model. The same
approach can also be used to validate the parameter estimates obtained from
Bayesian inference. Thus, we simulate 50 virtual cats using the means of the
posterior distributions of parameters for the SCDAs model. Figure
\ref{bedas_sim.eps} plots 20 paths. Comparison of the first four criteria
between the virtual cats and the observed cats results in the $P$ values of
0.074, 0.59, 0.34 and 0.41. The~last criterion, the die-out rate, of the simulated
cats is 14\%. While no die-outs were observed among the six observed cats,
this percentage is still less than 1$/$6.%

\begin{figure}

\includegraphics{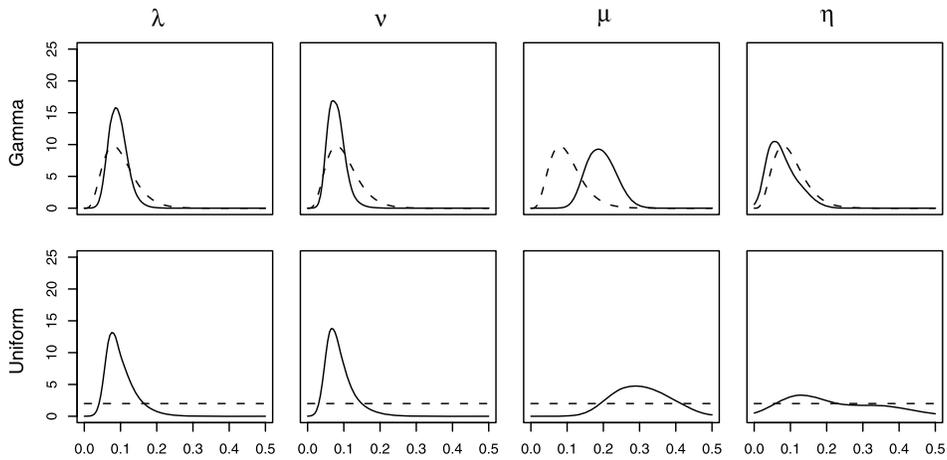}

\caption{Posterior distributions of the parameters for both SCD and SCDAs
models. Priors and posteriors are drawn in dashed and solid lines,
respectively. 100 weeks data. The prior is $\operatorname{Gamma}(5,50)$ for
the first row and $\operatorname{Uniform}(0,0.5)  $ for the second row.}%
\label{bedas_100wks_priors_pdist.eps}
\end{figure}

In this Bayesian analysis, we have to specify priors for the parameters.
Although there is some information guiding the choice of priors, we want to
check the sensitivity of the posterior distributions on prior choice. To that
end, we run MCMC with two priors. Both priors assume that all parameters are
independently and identically distributed. The ``Gamma'' priors let each
parameter be distributed as a Gamma distribution with shape parameter 5 and
rate parameter 50. The ``Uniform'' prior lets each parameter be distributed as
a uniform distribution between 0 and~0.5. While the SCD model is fairly
insensitive to the choice of priors [%
\citet{golinelli2006bih}%
], the SCDAs model is quite sensitive, particularly for clonal death rate and
asymmetric division rate (Figure~\ref{bedas_100wks_priors_pdist.eps}). For
these two rates, uniform priors lead to larger posterior means. Furthermore,
there appears to be little detailed information about asymmetric division
rate.\looseness=1

Both SCDAs and SCD are biologically reasonable models of hematopoiesis. Do the data at hand provide any evidence of favoring one versus the other?
The~Bayes factor comparing the SDAsAp model with the SCD model is $19$ (Table
\ref{TabIntLik}), providing moderate evidence that $\mathit{SCDAs}$ describes the data
better than $\mathit{SCD}$. Other than the Bayes factor approach, we could potentially
address this problem by a mixture model approach, treating the model indicator
$I$ as a random variable and putting a Bernoulli(0.5) prior on it. We try to
construct a reversible jump MCMC to sample the posterior joint distribution of
$I$, parameters and path. Unfortunately, due to the large dimensionality of
the path space, it seems very difficult to come up with a proposal
distribution for an RJMCMC algorithm that works.

\section{Homogeneity of the cats}\label{sec_homog}

\begin{figure}

\includegraphics{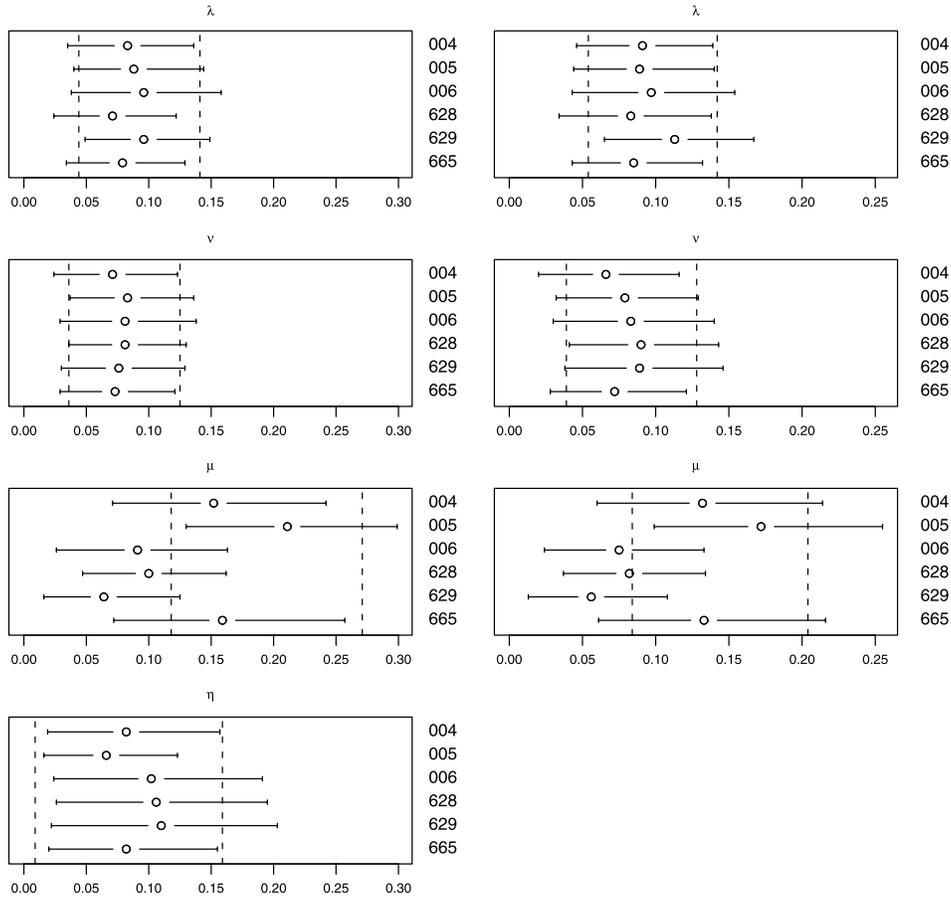}

\caption{95\% posterior credible intervals of the rate parameters for the SCDAs
(left) and SCD (right) model for individual cats. The circles in the middle of
the intervals mark the means of the posterior distributions. Dotted lines show
the 95\% posterior credible intervals for when all six cats are analyzed
together. The labels on the right of each panel are the identifiers of the
cats. Gamma prior, 100 weeks data.}
\label{single_cats_bed_bedas.eps}
\end{figure}

As we mentioned before,
\citet{golinelli2006bih}
showed that the posterior distributions of the rate parameters were fairly
insensitive to the choice of priors when all six cats were analyzed together.
\citet{golinelli2006bih}
also showed that when individual cats were analyzed separately, the posterior
distributions were quite sensitive to the choice of priors. This prompts us to
test the modeling assumption that the rate parameters are the same for all
cats. In order to test this assumption, we fit each cat separately with the
SCDAs model and the SCD model (Figure \ref{single_cats_bed_bedas.eps} left and
right) using the Gamma prior and compare the 95\% posterior credible intervals
of the rate parameters. It appears that both $\lambda$ and $\nu$ are quite
similar across cats for both the SCDAs model and the SCD model. In both
models, $\mu$ shows a bigger variation, but still all the credible intervals
for individual cats overlap with the credible interval for all cats together.
$\eta$ of the SCDAs model are also quite similar across cats.

Fitting each cat separately is equivalent to fitting a model in which each cat
is allowed to have its own set of parameters (the heterogeneous model). Viewed
this way, we can use the Bayes factor approach to compare pairs of models where
each pair differs in how each cat's parameters are treated. The Bayes factor
comparing the SCD model in which each cat is allowed to have its own set of
parameters and the SCD model in which all cats have the same set of parameters
is $19$. For the SCDAs model, this comparison yields a Bayes factor of
$0.059$. This tells us that in the smaller SCD model, the heterogeneous model
provides a better fit to the data, while in the bigger model SCDAs, the
heterogeneous model results in a worse fit.\looseness=1

\section{Discussion}\label{sec_discussion}

In this paper we use a number of statistical inference methods to answer some
important questions related to hematopoiesis. It is necessary for us to model
hematopoiesis as a continuous time Markov process because samples are
collected from unevenly-spaced time intervals. As a result, the sample space
that our Gibbs sampler has to explore is very large. We check the convergence
by visual inspection of the trace plots and by cusum plots [%
\citet{Yu1995}%
].

Previously
\citet{golinelli2006bih}
have shown that in the SCD model the amount of information the data contain
are different from different rate parameters. More information is available
for the symmetric division rate and the commitment rate than for the clonal
death rate. We show here that in the SCDAs model, the same is true.
Furthermore, the amount of information for the asymmetric division rate is similar
to that for the clonal death rate. In other words, these two parameters are
not estimated very precisely.

The data we analyze in this paper are observations made up to 100 weeks
post-transplantation. Data from 100 weeks to 300 weeks post-transplantation
are also available. We choose to limit the analysis to 100 weeks data out of
consideration of the underlying physiological process. In the initial period
after transplantation, the number of HSC cells grow exponentially until
reaching the number of niches available. After that, following a symmetric
division event, one of the daughter cells may not have a niche available to it
and end up dying [%
\citet{schofield1978rbs}%
] or specializing (differentiating). We call this the steady state. In the
steady state, the number of HSC cells is bounded by the number of niches
available and the amount of information for the rate parameters is limited.
Since this mechanism operates even in the SCD model, there is in effect a
modest amount of asymmetric division even in this case.

The model has been applied to other animals [%
\citet{abkowitz2000vkm}%
,
\citet{shepherd2007hsc}%
] with results that indicate that the kinetics of\break hematopoiesis varies
substantially between animals. There are some surprising invariants between
species, such as an indication that the maximum number of stem cells is
similar for all animals, that the ratio of $\lambda$ to $\nu$ remains constant
[%
\citet{abkowitz2002enh}%
], and that there is no evidence that the clone exhaustion parameter $\mu$
differs between species, although our results indicate substantial uncertainty
about this latter parameter. These findings have been used to deduce parameter
values for humans [%
\citet{shepherd2004}%
, \citet{catlinsubmitted}].

There is independent evidence that the parameter values obtained in our
analyses are reasonable. Specifically, observations of telomere shortening in
granulocytes from cats with aging can be used to estimate the cumulative
numbers of HSC divisions at a point in time [Shepherd et al. (\citeyear{shepherd2007hsc})]. While the
estimates in this paper are based on a stochastic model using only symmetric
division, similar calculations indicate that even though the fitted values for
the SCDAs model will be about 185 divisions per lifetime, it is quite consistent
with the other animal models in the cited paper.

Recently there has been progress in studies of in vitro asymmetric fate
determination.
\citet{wu2007ihp}
for the first time demonstrated that hematopoietic stem/progenitor cells
undergo both symmetric and asymmetric divisions. The~study we have done in
this paper is consistent with the occurrence of asymmetric division in vivo
and further provides an estimate of the rate at which it happens.

\begin{appendix}\label{append}
\section*{Appendix}

The state update is carried out using a Metropolis--Hastings--Green algorithm
which is a slight generalization and optimization of that reported in
\citet{golinelli2006bih}%
. In this appendix we only lay out what is different and ask the readers to
refer to
\citet{golinelli2006bih}
for more details. Denote by $M$ the types of events in the model we are
interested in, by $N$ the number of events in the current path, and by $[
0,T]  $ the time span during which data are collected. Let $(
p_{1},p_{2},p_{3})  $ be a point from the probability simplex. The
proposal probabilities for the three moves defined in
\citet{golinelli2006bih}
are as follows:

\begin{enumerate}
\item \textit{Deletion move}: $p_{1}/N$;

\item \textit{Insertion move}: $p_{2}/(  2MT)  $;

\item \textit{Shuffle move}: $p_{3}/(  NT)  $.
\end{enumerate}

In calculating the acceptance probability
\[
R=\frac{p(  \omega^{\prime}|\theta)  }{p(  \omega
|\theta)  }\frac{p(  \mathbf{y}|\omega^{\prime})  }{p(
\mathbf{y}|\omega)  }\frac{q(  \omega|\omega^{\prime})
}{q(  \omega^{\prime}|\omega)  }\vert J\vert ,
\]
the Jacobian $\vert J\vert $ is 1 for all models investigated in
this paper since the diffeomorphisms are all identity relations. For speed and
ease of computation, we factor the log prior ratio $\log\{  p(
\omega^{\prime}|\theta)  /p(  \omega|\theta)  \}  $
into three parts:

\begin{itemize}
\item The first part is $\log(  \prod_{k\in\mathcal{\bar{D}}}%
z_{k-1}^{\prime}\prod_{k\in\mathcal{D}}x_{k-1}^{\prime})
-\log(  \prod_{k\in\mathcal{\bar{D}}}z_{k-1}\prod_{k\in\mathcal{D}%
}x_{k-1})  $, where $z_{k-1}$ denotes the number of cells in the first
compartment right before the $k$th event and $x_{k-1}$ denotes the number
of clones in the second compartment right before the $k$th event.
$k\in\mathcal{\bar{D}}$ if and only if the $k$th event is not a Death event.

\item The second part is $\log(  \lambda^{S^{\prime}-S}\nu^{C^{\prime}%
-C}\mu^{D^{\prime}-D}\eta^{F^{\prime}-F}\alpha^{A^{\prime}-A})  $, where
$S$, $C$, $D$, $F$ and~$A$ are the numbers of symmetric division, commitment,
clonal death, asymmetric division and apoptosis events, separately. When a
model lacks a certain event, simply drop the corresponding term. After
simplification, this part is just $\delta\log\rho$, where $\delta=1$, $-1$ or
$0$ for insertion, deletion and shuffle moves, respectively, and $\rho$ is the
rate parameter corresponding to the event being considered in the proposal.

\item The third part is $-(  \lambda+\nu+\eta+\alpha)  (
S^{\prime z}-S^{z})  -\mu(  S^{\prime x}-S^{x})  $, where
$S^{z}$ and $S^{x}$ are the total time lived by the population in the first
and second compartment, respectively. Again, when a model lacks a certain
event, simply drop the corresponding term.
\end{itemize}
The likelihood ratio $\log\{  p(  \mathbf{y}%
|\omega^{\prime})  /p(  \mathbf{y}|\omega)  \}  $ is
calculated similarly for all models. The proposal ratio $\log\{  q(
\omega|\omega^{\prime})  /q(  \omega^{\prime}|\omega)
\}  $ is $\log(  \frac{p_{1}}{p_{2}}\frac{2MT}{N+1})  $ for
the insertion move,\vspace*{-1pt} $\log(  \frac{p_{2}}{p_{1}}\frac{N}{2MT})  $
for the deletion move, and 0 for the shuffle move.
\end{appendix}

\section*{Acknowledgments}

The authors thank the Editor, Associate Editor and referees for helpful comments.

\printaddresses

\end{document}